\documentclass[journal,11pt,draftclsnofoot,onecolumn]{IEEEtran}

\usepackage{graphicx}

\usepackage[sort,compress]{cite}

\usepackage{latexsym}
\usepackage{psfrag}
\usepackage{amsthm,amssymb,amscd,amssymb,amsmath,mathrsfs}
\usepackage[noend]{algpseudocode}
\usepackage{color}

\theoremstyle{remark}

\newtheorem{lem}{Lemma}
\newtheorem{thm}{Theorem}
\newtheorem{defn}{Definition}

\begin{document}

\title{ {\huge Average Consensus on Arbitrary Strongly Connected Digraphs with Time-Varying Topologies} }

\author{ Kai Cai and Hideaki Ishii 

\thanks{Kai Cai is with Department of Electrical and Computer
Engineering, University of Toronto, Toronto, ON, M5S 3G4, Canada.
kai.cai@scg.utoronto.ca. Hideaki Ishii is with Department of
Computational Intelligence and Systems Science, Tokyo Institute of
Technology, Yokohama 226-8502, Japan. ishii@dis.titech.ac.jp. This
work was supported in part by the Ministry of Education, Culture,
Sports, Science and Technology in Japan under Grants-in-Aid
for Scientific Research, No.\ 21760323 and 23760385.}
}

\maketitle

\begin{abstract}
We have recently proposed a ``surplus-based'' algorithm which solves
the multi-agent average consensus problem on general strongly
connected and static digraphs.  The essence of that algorithm is to
employ an additional variable to keep track of the state changes of
each agent, thereby achieving averaging even though the state sum is
not preserved.  In this note, we extend this approach to the more
interesting and challenging case of time-varying topologies: An
extended surplus-based averaging algorithm is designed, under which
a necessary and sufficient graphical condition is derived that
guarantees state averaging. The derived condition requires only that
the digraphs be arbitrary strongly connected in a \emph{joint}
sense, and does not impose ``balanced'' or ``symmetric'' properties
on the network topology, which is therefore more general than those
previously reported in the literature.
\end{abstract}




\section{Introduction} \label{Sec_Intro}

The average consensus problem of multi-agent systems has attracted
much attention in the literature (e.g.,
\cite{BerTsi:89,OlfMur:04,XiBo:04}).  The problem can be described
as follows.  Consider a network of $n$ agents whose state is $x(k) =
[x_1(k) \ \cdots \ x_n(k)]^T \in \mathbb{R}^n$ at discrete time
$k=0,1,2,\ldots$.  Every agent $i \in [1,n]$ interacts locally with
its neighbors for the exchange of state information, and based on
the obtained neighbors' states it updates its own $x_i(k)$ to a new
value $x_i(k+1)$ according to a prescribed algorithm.  One aims at
designing distributed algorithms by which agents may iteratively
update their states such that $x(k)=x_a\textbf{1}$ asymptotically,
where $x_a:=\textbf{1}^Tx(0)/n$ is the average of the initial states
and $\textbf{1}:=[1 \ \cdots \ 1]^T \in \mathbb{R}^n$.

In \cite{CaiIshii_Aut:12} we proposed a novel algorithm which
provably achieves average consensus on \emph{general} strongly
connected, static networks.  This result extends
\cite{OlfMur:04,XiBo:04} in that it does not require the
``balanced'' property on the network topology which can be
restrictive as every agent needs to maintain exactly equal amounts
for incoming and outgoing information.  This is realized by
augmenting for each agent an additional variable $s_i \in
\mathbb{R}$, which we call ``surplus''.  Each surplus $s_i(k)$ at
time $k$ keeps track of the state change $x_i(k)-x_i(k-1)$ of agent
$i$, in such a way that $\textbf{1}^T (x(k)+s(k))$ is time-invariant
(here $s(k) = [s_1(k) \ \cdots \ s_n(k)]^T$) despite that the state
sum $\textbf{1}^Tx(k)$ is in general not.  The idea was originated
in \cite{CaiIshii_ACC:10} for dealing with a quantized averaging
problem.

A more interesting, yet more challenging, scenario is where the
agents' network topology is dynamic, as opposed to static.  In real
networks, many practical factors could result in a dynamic topology.
There can be unpredictable communication issues like random packet
loss, link failure, and node malfunction. There might also exist
deterministic, supervisory switchings among different modes of the
network.  A gossip-type randomized dynamic topology has been
considered in \cite{CaiIshii_Aut:12}, where we proved that an
arbitrary strongly connected topology in \emph{expectation} is
necessary and sufficient for our surplus-based algorithm to achieve
average consensus in \emph{mean-square} and \emph{almost surely}. In
this note, we focus on dynamic network topology varying in some
deterministic fashion, and design an extended surplus-based
algorithm to achieve state averaging in a uniform sense (defined
below). Parts of the results here are contained in the conference
precursor \cite{CaiIshii_ACC:12}.

Our main contribution is that the required connectivity condition on
time-varying network topology is weakened, as compared to those
previously reported in the literature. In \cite{OlfMur:04}, it was
shown that a sufficient connectivity condition for average consensus
is that the network topology at every time (possibly different)
should be both strongly connected and balanced. By contrast,
supported by surplus variables, we justify that average consensus
can be uniformly achieved if and only if the dynamic network is
\emph{jointly} strongly connected (the precise definition is given
in Section~\ref{Sec_ProFor}). Thus for one, the ``balanced''
requirement at every instant is dropped; for the other, ``strongly
connected'' is needed only in a joint sense. As to the convergence
proof, we use a Lyapunov-type argument, in the spirit of
\cite{Mor:05}.  Extending the algorithm in \cite{CaiIshii_Aut:12},
we introduce a new switching mechanism, which gives rise to a
suitable Lyapunov function for state evolution. Finally, when the
derived result is specialized to the static network case, we
effectively relax a conservative requirement on a parameter of the
algorithm in \cite{CaiIshii_Aut:12}.

{\color{black} There are well-known results (existence of a spanning
tree jointly, e.g., \cite{Mor:05,Lin:Mono:08}) for achieving a
general consensus over dynamic networks, as well as new conditions
of cut-balanced in \cite{HenTsi:13}.  To further achieve the special
average consensus on the initial state, either the state sum is kept
invariant or there is a way of tracking the changes of the state
sum.  We consider arbitrary strongly connected dynamic topologies
where the state sum is time-varying in general, and propose
additional surplus update dynamics to keep track of the state
changes of individual agents. The surplus values are used in turn to
influence the state update dynamics, thereby forcing the states to
converge to, and only to, the initial average value.}

We note that \cite{FraGiuSea:11,BBTTV:10} also addressed average
consensus on general dynamic networks by employing auxiliary
variables. In \cite{FraGiuSea:11}, an auxiliary variable is
associated to each agent and a linear ``broadcast gossip'' algorithm
is proposed; however, the convergence of that algorithm is not
proved. Reference \cite{BBTTV:10} also uses extra variables, and a
nonlinear (division involved) algorithm is designed and proved to
achieve state averaging on non-balanced digraphs.  The idea is based
on computing the stationary distribution for the Markov chain
characterized by the agent network, and is thus different from
consensus-type algorithms \cite{BerTsi:89,OlfMur:04,XiBo:04}.
Moreover, the dynamic networks considered are of randomized type;
consequently the algorithms and results are not directly applicable
to the deterministic time-varying case studied in this note.
{\color{black} In addition, \cite{DomHad:13} presents distributed
algorithms which iteratively update a column-stochastic matrix into
a doubly-stochastic one, and then embeds this matrix update into a
standard consensus state update to achieve average consensus. Since
the time-varying update matrices used are all column-stochastic, the
state sum is invariant in \cite{DomHad:13}; this is different from
the case of time-varying state sum we study here.}  Finally,
centralized and distributed algorithms are designed in
\cite{GhaCor:11} to make a general static topology balanced. The
algorithm may in principle be used also for dynamic networks, which
would require a complete execution at each time for different
topologies. This requirement might be strong for applications where
networks vary fast.




The rest of the paper is organized as follows. First, in
Section~\ref{Sec_ProFor} we formulate the average consensus problem
for deterministic time-varying networks.  Then an extended
surplus-based algorithm is designed in Section~\ref{Sec_Det}, and
the corresponding convergence result presented and proved in
Section~\ref{Sec_Prf}. A numerical example is shown in
Section~\ref{Sec_Simul}, and finally in Section~\ref{Sec_Concl} we
state our conclusions.


\section{Average Consensus Problem} \label{Sec_ProFor}

First, a review of graph notions relevant to this note is provided;
and then, the average consensus problem on deterministic
time-varying networks is formulated.

For a network of $n$ agents, we model their time-varying
interconnection structure at time $k$ by a \emph{dynamic digraph}
$\mathcal {G}(k) = (\mathcal {V}, \mathcal {E}(k))$: Each node in
$\mathcal {V} = \{ 1,...,n \}$ stands for an agent, and each
directed edge $(j,i)$ in $\mathcal {E}(k) \subseteq \mathcal {V}
\times \mathcal {V}$ represents that agent $j$ communicates to agent
$i$ at time $k$.  For each node $i \in \mathcal {V}$, let $\mathcal
{N}_i^+(k):=\{j \in \mathcal {V}:(j,i)\in \mathcal {E}(k)\}$ denote
the set of its ``in-neighbors'', and $\mathcal {N}_i^-(k):=\{j \in
\mathcal{V}:(i,j)\in \mathcal {E}(k)\}$ the set of its
``out-neighbors''.  Also we adopt the convention $(i,i) \notin
\mathcal {E}(k)$ and $i \notin \mathcal{N}_i^+(k), \mathcal
{N}_i^-(k)$.

For the dynamic digraph $\mathcal {G}(k)$, we introduce a notion of
\emph{joint connectivity} over some finite time interval.  In
$\mathcal {G}(k)$ a node $i$ is \emph{reachable} from a node $j$ if
there exists a sequence of directed edges from $j$ to $i$ which
respects the direction of the edges.  We say $\mathcal {G}(k)$ is
\emph{strongly connected} if every node is reachable from every
other node.  For a time interval $[k_1,k_2]$ define the \emph{union
digraph} $\mathcal{G}([k_1,k_2]) := \left( \mathcal {V}, \bigcup_{k
\in [k_1,k_2]} \mathcal {E}(k) \right)$; namely, the edge set of
$\mathcal {G}([k_1,k_2])$ is the union of those over the interval
$[k_1,k_2]$.  A dynamic digraph $\mathcal {G}(k)$ is \emph{jointly
strongly connected} if there is $k_1$ such that for every $k_0$ the
union digraph $\mathcal {G}([k_0,k_0+k_1])$ is strongly connected.

The ``joint'' type connectivity notions have appeared in many
previous works, e.g., \cite{Mor:05,RenBea:08,Lin:Mono:08}. In
particular, to achieve a general consensus (where the consensus
value need not be the initial average $x_a$), the following joint
connectivity is essential. A node $v \in \mathcal {V}$ is called a
\emph{globally reachable node} if every other node is reachable from
$v$. A dynamic digraph $\mathcal {G}(k)$ \emph{jointly contains a
globally reachable node (or a spanning tree)} if there is $k_1$ such
that for every $k_0$ the union digraph $\mathcal {G}([k_0,k_0+k_1])$
contains a globally reachable node. It is shown in
\cite{Mor:05,RenBea:08,Lin:Mono:08} that a general consensus can be
uniformly achieved on a dynamic digraph $\mathcal {G}(k)$ if and
only if $\mathcal {G}(k)$ jointly contains a globally reachable
node. This joint connectivity notion is weaker than the above
``jointly strongly connected'' notion, because a strongly connected
union digraph $\mathcal {G}([k_0,k_0+k_1])$ is equivalent to that
every node of $\mathcal {G}([k_0,k_0+k_1])$ is globally reachable.
This notion is, however, too weak to achieve average consensus, as
we will see in the necessity proof of our main result; there we show
that the ``jointly strongly connected'' notion is, indeed, a
necessary and sufficient condition for uniformly achieving average
consensus.

We present several additional graph notions, which will be needed in
the necessity proof of our main result. For $\mathcal {G}(k) =
(\mathcal {V}, \mathcal {E}(k))$ and a nonempty subset $\mathcal
{U}$ of $\mathcal {V}$, we say $\mathcal {U}$ is \emph{closed} if
every node $u$ in $\mathcal {U}$ is not reachable from any node $v$
in $\mathcal {V} - \mathcal {U}$ at time $k$. Also, the digraph
$\mathcal {G}(k)_\mathcal {U} = (\mathcal {U}, \mathcal {E}(k) \cap
(\mathcal {U} \times \mathcal {U}))$ is called the \emph{induced
subdigraph} by $\mathcal {U}$.   Lastly, a \emph{strong component}
of $\mathcal {G}(k)$ is a maximal induced subdigraph of $\mathcal
{G}(k)$ which is strongly connected.

The average consensus problem on deterministic time-varying networks
is formulated as follows.


\begin{defn} \label{defn:det}
A network of agents achieves \emph{uniform average consensus} if for
all $c_1,c_2>0$ there exists $k_1$ such that for every $k_0$,
\begin{align*}
|| (x(k_0),s(k_0)) - (x_a\textbf{1}, 0) ||_\infty < c_1 \
\Rightarrow \ (\forall k \geq k_0+k_1)\ || (x(k),s(k)) -
(x_a\textbf{1}, 0) ||_\infty < c_2.
\end{align*}
\end{defn}

The above definition of average consensus is in a ``uniform'' sense
with respect to $k_0$. For studying consensus on deterministic
time-varying networks, this uniform consensus notion is typical,
e.g., \cite{Mor:05,Lin:Mono:08}.

\emph{Problem:} Design {\color{black} a distributed algorithm} and
find a necessary and sufficient connectivity condition on dynamic
digraphs such that the agents achieve uniform average consensus.


\section{Surplus-Based Averaging Algorithm} \label{Sec_Det}

In this section, we present a surplus-based averaging algorithm,
which is an extension of the one in \cite{CaiIshii_Aut:12}.
Implementation issues of the algorithm are discussed, and basic
properties of the algorithm are shown.

In the algorithm, there are three operations that every agent $i$
performs at time $k$.  First (sending stage), agent $i$ sends its
state $x_i(k)$ and weighted surplus $b_{ih}(k) s_i(k)$ to each
out-neighbor $h \in \mathcal {N}_i^-(k)$ {\color{black} (weights
$b_{ih}(k)$ are specified below)}. Second (receiving stage), agent
$i$ receives state $x_j(k)$ and weighted surplus $b_{ji}(k) s_j(k)$
from each in-neighbor $j \in \mathcal {N}_i^+(k)$. Third (updating
stage), agent $i$ updates its own state $x_i(k)$ and surplus
$s_i(k)$ as follows:
\begin{equation}\label{eq:state}
\begin{split}
x_i(k+1)= x_i(k) + c_i(k)\sum_{j \in \mathcal {N}_i^+(k)}a_{ij}(k)
(x_j(k)-x_i(k)) + \epsilon_i(k) s_i(k)
\end{split}
\end{equation}
\begin{equation}\label{eq:surplus}
\begin{split}
s_i(k+1) = ( 1-\sum_{h \in \mathcal {N}_i^-(k)} b_{ih}(k) ) s_i(k) +
\sum_{j \in \mathcal {N}_i^+(k)} b_{ji}(k) s_j(k) -
\Big(x_i(k+1)-x_i(k)\Big)
\end{split}
\end{equation}
{\color{black} where the parameters $\epsilon_i(k), a_{ij}(k),
b_{ih}(k), c_i(k)$ used in (\ref{eq:state}) and (\ref{eq:surplus})
satisfy the following items, for every $i,j,h \in \mathcal {V}$ and
every $k$:}
\begin{description}
  \item[(P1)] The parameter $\epsilon_i(k) \in (0,1)$, which specifies the amount of surplus used for state
  update.
  \item[(P2)] The updating weights $a_{ij}(k) \in (0,1)$ if $j \in \mathcal {N}_i^+(k)$, $a_{ij}(k)=0$ otherwise, and $\sum_{j \in \mathcal{N}_i^+(k)}
  a_{ij}(k)<1$.
  \item[(P3)] The sending weights $b_{ih}(k) \in (0,1)$ if $h \in \mathcal {N}_i^-(k)$, $b_{ih}(k)=0$ otherwise, and $\sum_{h \in \mathcal
  {N}_i^-(k)}$ $b_{ih}(k)<1-\epsilon_i(k)$. The last inequality means that the
  amount of surplus sent to out-neighbors should be strictly less
  than {\color{black} the total surplus subtracted by the part used for state
  update.}
  \item[(P4)] The switching parameters $c_i(k) = 1$ if $\sum_{j \in \mathcal {N}_i^+(k)} a_{ij}(k) (x_j(k)-x_i(k)) \leq 0$, and $c_i(k) = 0$
  otherwise. This means that whenever an agent
  determines to make a \emph{positive} state update based on the
  information from in-neighbors, it may use only its
  surplus for that update.
\end{description}
(P1)-(P4) will enable desired properties of the proposed algorithm.
In particular, (P3) and (P4) will establish that all the surpluses
are \emph{nonnegative}; see Lemma~\ref{lem:property} below. Note
also that at the sending stage of the algorithm, each agent should
know its out-neighbors at time $k$, namely the members of $\mathcal
{N}_i^-(k)$.

We discuss the implementation of the above protocol in applications
of sensor networks. Let $\mathcal {G}(k) = (\mathcal {V}, \mathcal
{E}(k))$ represent a dynamic network of sensor nodes.  Our protocol
deals particularly with scenarios where information flow among
sensors is directed and time-varying. A concrete example is using
sensor networks for monitoring geological areas (e.g., volcanic
activities), where sensors are fixed at certain locations. At the
time of setting them up, the sensors may be given different
transmission power for saving energy (such sensors must run for a
long time) or owing to geological reasons. Once the power is fixed,
the neighbors (and their IDs) can be known to each sensor; at time
$k$, each sensor may choose to broadcast its information to all
neighbors, or to communicate with a random subset of neighbors, or
even not to communicate at all (saving power). Thus, a directed and
time-varying topology can arise in this sensor networks application.
To implement states and surpluses, we see from (\ref{eq:state}),
(\ref{eq:surplus}) that they are ordinary variables locally stored,
updated, and exchanged; thus they may be implemented by allocating
memories in sensors. Similarly, since the values of the time-varying
weights $a_{ij}(k)$, $b_{ih}(k)$ and parameters $c_i(k)$,
$\epsilon_i(k)$ can all be locally determined, these variables may
be implemented as sensors' memories as well.

Now define the \emph{adjacency matrix} $A(k)$ of the digraph
$\mathcal {G}(k)$ by $A(k):=[c_i(k) a_{ij}(k)]$. Then the
\emph{Laplacian matrix} $L(k)$ is defined as $L(k):=D(k)-A(k)$,
where $D(k)=\mbox{diag}(d_1(k), \ldots, d_n(k))$ with
$d_i(k)=\sum_{j=1}^n c_i(k) a_{ij}(k)$.  It is easy to see that
$L(k)$ has nonnegative diagonals, nonpositive off-diagonal entries,
and zero row sums. Consequently the matrix $I-L(k)$ is nonnegative
(by $\sum_{j \in \mathcal {N}_i^+(k)}a_{ij}(k)<1$ in (P2)), and
every row sums up to one; namely $I-L(k)$ is \emph{row stochastic}.

Also, let $B(k):=[b_{ih}(k)]^T$ (note that the transpose in the
notation is needed because $h \in \mathcal {N}_i^-(k)$ for
$b_{ih}(k)$).  Define the matrix $S(k):=(I-\tilde{D}(k))+B(k)$,
where $\tilde{D}(k)=\mbox{diag}(\tilde{d}_1(k), \ldots,$
$\tilde{d}_n(k))$ with $\tilde{d}_i(k)=\sum_{h=1}^n b_{ih}(k)$. Then
$S(k)$ is nonnegative (by $\sum_{h \in \mathcal {N}_i^-(k)}
b_{ih}<1-\epsilon_i(k)$ in (P3) and $\epsilon_i(k) \in (0,1)$ in
(P1)), and every column sums up to one; that is, $S(k)$ is
\emph{column stochastic}. As can be observed from
(\ref{eq:surplus}), $S(k)$ captures the part of the update induced
by sending and receiving surpluses.  Finally, let $E(k) :=
\mbox{diag}(\epsilon_1(k), \ldots, \epsilon_n(k))$.

With the above matrices defined, the iteration of states
(\ref{eq:state}) and surpluses (\ref{eq:surplus}) can be
written in the following matrix form:
\begin{equation} \label{eq:dyn_alg}
\begin{split}
\hspace{0cm} \begin{bmatrix}
  x(k+1)\\
  s(k+1)
\end{bmatrix}=
M(k)
\begin{bmatrix}
  x(k)\\
  s(k)
\end{bmatrix},
\mbox{where } M(k):=
\begin{bmatrix}
  I-L(k) & E(k)\\
  L(k) & S(k)-E(k)
\end{bmatrix} \in \mathbb{R}^{2n \times 2n}.
\end{split}
\end{equation}
Notice that the matrix $M(k)$ has negative entries due to the
presence of the Laplacian matrix $L(k)$ in the $(2,1)$-block. Note
also that the column sums of $M(k)$ are equal to one (here $S(k)$
being column stochastic is crucial), which implies that the quantity
$\textbf{1}^T (x(k)+s(k))$ is a constant for all $k$.

Some other useful implications derived from this
algorithm~(\ref{eq:dyn_alg}) are collected in the following lemma.
Define the minimum and maximum states, $\underline{m}(x)$ and
$\overline{m}(x)$, respectively, by
\begin{align} \label{eq:minmax}
\underline{m}(x) := \min_{i \in \mathcal {V}} x_i,\ \ \
\overline{m}(x) := \max_{i \in \mathcal {V}} x_i.
\end{align}

\begin{lem} \label{lem:property}
In the algorithm~(\ref{eq:dyn_alg}), the following properties hold:
\begin{description}
  \item[(i)] The surplus is nonnegative, $s_i(k) \geq 0$, for every $i \in \mathcal {V}$ and $k$.
  \item[(ii)] The minimum state $\underline{m}(x)$ is non-decreasing, i.e., $\underline{m}(x(k_1)) \leq \underline{m}(x(k_2))$ if $k_1 \leq k_2$.
  \item[(iii)] The minimum state satisfies $\underline{m}(x(k)) \leq x_a$ for every $k \in
  \mathbb{Z}_+$; and $\underline{m}(x(k)) = x_a$ implies $(\forall i \in \mathcal {V})\ x_i(k)=x_a$ and
  $s_i(k)=0$, i.e., average consensus.
  {\color{black} \item[(iv)] The unique equilibrium of (\ref{eq:dyn_alg}) is
  $(x_a\textbf{1},0)$.}
\end{description}
\end{lem}
\emph{Proof.} (i) We show this property by
induction on the time index $k$. For the base case $k=0$, we have
$s_i(0)=0$ for all $i$.  Now suppose that $s_i(k) \geq 0$, $k>0$,
for all $i$. According to (\ref{eq:state}) and (\ref{eq:surplus})
we derive
\begin{align*}
s_i(k+1) = &\Big( 1-\sum_{h \in \mathcal {N}_i^-(k)} b_{ih}(k)-\epsilon_i(k) \Big)s_i(k)\\
& + \sum_{j \in \mathcal {N}_i^+(k)} b_{ji}(k) s_j(k) - \sum_{j \in \mathcal {N}_i^+(k)}c_i(k) a_{ij}(k) (x_j(k)-x_i(k)).
\end{align*}
It then follows from (P3), (P4), and the induction
hypothesis that $s_i(k+1) \geq 0$ for all $i$.  This completes the induction.

(ii) Let $k$ be arbitrary. First consider a node $i \in \mathcal
{V}$ such that $x_i(k)=\underline{m}(x(k))$. It must hold that
$\sum_{j \in \mathcal {N}_i^+(k)} a_{ij}(k) (x_j(k)-x_i(k)) \geq 0$.
Thus by (\ref{eq:state}) and (P4), the state update of node $i$ is
$x_i(k+1) = x_i(k)+\epsilon_i(k) s_i(k) \geq
x_i(k)=\underline{m}(x(k))$. {\color{black} Next consider a node $i$
such that $x_i(k)>\underline{m}(x(k))$; there are two cases. Case 1:
$c_i(k)=0$. Then $x_i(k+1) = x_i(k)+\epsilon_i(k) s_i(k) \geq
x_i(k)>\underline{m}(x(k))$. Case~2: $c_i(k)=1$. Then $x_i(k+1) =
x_i(k)+\sum_{j \in \mathcal {N}_i^+(k)} a_{ij}(k)
(x_j(k)-x_i(k))+\epsilon_i(k) s_i(k)$. Notice that the first two
terms of the above summation consist of a convex combination of
$x_i(k)$ and $x_j(k)$, $j\in \mathcal {N}_i^+(k)$, and hence
$x_i(k)+\sum_{j \in \mathcal {N}_i^+(k)} a_{ij}(k) (x_j(k)-x_i(k)) >
\min_{j \in \{i\} \cup \mathcal {N}_i^+(k) } x_j(k) \geq
\underline{m}(x(k))$. In turn $x_i(k+1)>\underline{m}(x(k))$.
Therefore, the minimum state cannot decrease.}

(iii) Suppose on the contrary that $\underline{m}(x(k)) > x_a$ for
some $k$. This implies that $\textbf{1}^Tx(k)+\textbf{1}^Ts(k)>n x_a
+ \textbf{1}^Ts(k)$.  But since
$\textbf{1}^Tx(k)+\textbf{1}^Ts(k)=\textbf{1}^Tx(0)=n x_a$, one
obtains $\textbf{1}^Ts(k)<0$, a contradiction to the property (i).
Hence we conclude that $\underline{m}(x(k)) \leq x_a$ for all $k$.
And when $\underline{m}(x(k))=x_a$, we must also have
$\overline{m}(x(k))=x_a$ owing again to (i). Therefore $x_i(k)=x_a$
and $s_i(k)=0$ for all $i$.

{\color{black} (iv) For every $i \in \mathcal {V}$, substituting
$x_i(k)=x_a$ and $s_i(k)=0$ into equations (\ref{eq:state}) and
(\ref{eq:surplus}) yields $x_i(k+1)=x_i(k)$ and $s_i(k+1)=s_i(k)$.
Hence $(x_a\textbf{1},0)$ is an equilibrium of (\ref{eq:dyn_alg}).
For uniqueness, suppose $(x,s) \neq (x_a\textbf{1},0)$ is another
equilibrium. Then by $x_i(k+1)=x_i(k)$ in (\ref{eq:state}) we have
$c_i(k)\sum_{j \in \mathcal {N}_i^+(k)}a_{ij}(k) (x_j(k)-x_i(k)) +
\epsilon_i(k) s_i(k) = 0$ for all $i$.  Since $s_i(k) \geq 0$
according to (i), it must hold that $s_i(k)=0$ and $x_i(k)=x_j(k)$,
for all $i,j \in \mathcal {V}$. So $(x,s)$ is of the form
$(x_b\textbf{1},0)$, $x_b \neq x_a$ (otherwise $(x,s) =
(x_a\textbf{1},0)$). However, $\textbf{1}^T(x+s)=nx_b \neq nx_a
=\textbf{1}^T(x(0)+s(0))$; this contradicts that $\textbf{1}^T
(x(k)+s(k))$ is a time-invariant quantity for the
algorithm~(\ref{eq:dyn_alg}). \hfill $\blacksquare$}


\section{Convergence Result and Proof} \label{Sec_Prf}

In this section, we present our main result and provide its proof.

\begin{thm} \label{thm:dyn_alg}
Using the algorithm~(\ref{eq:dyn_alg}), a network of agents achieves
uniform average consensus if and only if the dynamic digraph
$\mathcal {G}(k)$ is jointly strongly connected.
\end{thm}

Comparing our derived graphical condition with the one in
\cite{OlfMur:04}, we drop the balanced requirement at every moment
on one hand, and need strongly connected property only in a joint
sense on the other hand. Also, for the special case of static
digraphs, we can use the algorithm~(\ref{eq:dyn_alg}) with a fixed
constant parameter $\epsilon \in (0,1)$; there will still be
switching in the updates. However, the original algorithm in
\cite{CaiIshii_Aut:12} may not converge because this $\epsilon$
value might be too large for the algorithm to remain stable (in
\cite{CaiIshii_Aut:12}, $\epsilon$ is required to be
\emph{sufficiently small} (conservative bounds available) to ensure
convergence of the designed algorithms). Finally, the proof
techniques in \cite{CaiIshii_Aut:12} and here are very different:
\cite{CaiIshii_Aut:12} relied on matrix perturbation theory, while
here a Lyapunov-type argument is used, below.

We note that there have been efforts in the literature addressing
time-varying consensus/averaging problems with second order
dynamics. In \cite{LiuAndCaoMor:09}, an ``accelerated gossip''
algorithm is designed which relies heavily on symmetry of undirected
graphs. The algorithm studied in \cite{RenBea:08}, on the other
hand, is based on the assumption of dwell-time switching of the
time-varying topology. By contrast, we study general dynamic
digraphs that vary at every discrete time instant and each resulting
update matrix (\ref{eq:dyn_alg}) is not nonnegative.

We now proceed to the proof of Theorem~\ref{thm:dyn_alg}, for which
we rely on the following Lyapunov result (cf. \cite[Theorem~4 and
Remark~5]{Mor:05}). For any given $x_a$, let
\begin{align} \label{eq:statespace}
\mathcal{X}(x_a):=\{(x,s) : \textbf{1}^T (x+s)/n=x_a,\ s \geq 0\}.
\end{align}

\begin{lem} \label{lem:moreau_lyapunov}
Consider the algorithm~(\ref{eq:dyn_alg}). Suppose that continuous
functions $V:\mathcal{X}(x_a) \rightarrow \mathbb{R}_+$ and
$\delta:\mathcal{X}(x_a) \rightarrow \mathbb{R}_+$ satisfy the
following conditions:

\noindent (i) $V$ is bounded on bounded subsets of
$\mathcal{X}(x_a)$, and positive definite with respect to the
average consensus point $(x_a\textbf{1},0)$ (i.e.,
$V(x_a\textbf{1},0)=0$ and $V(x,s)>0$ if $(x,s) \neq
(x_a\textbf{1},0)$);

\noindent (ii) $\delta$ is also positive definite with respect to
the average consensus point $(x_a\textbf{1},0)$ (i.e.,
$\delta(x_a\textbf{1},0)=0$ and $\delta(x,s)>0$ if $(x,s) \neq
(x_a\textbf{1},0)$);

\noindent (iii) there exists a finite time $\kappa$ such that for
every $(x(k),s(k)) \in \mathcal{X}(x_a)$,
\begin{align*}
V(x(k+\kappa),s(k+\kappa)) - V(x(k),s(k)) \leq -\delta(x(k),s(k)).
\end{align*}
Then, the network of agents achieves uniform average consensus.
\end{lem}

Lemma~\ref{lem:moreau_lyapunov} is an application of the more
general result \cite[Theorem~4 and Remark~5]{Mor:05} to the dynamic
system~(\ref{eq:dyn_alg}) with the equilibrium $(x_a\textbf{1},0)$.
Note that the function $V$ in Lemma~\ref{lem:moreau_lyapunov}
corresponds to a compound $\mu \circ V'$ of two functions $V'$ and
$\mu$ in \cite[Theorem~4]{Mor:05}, where $V'$ is a set-valued
function on $\mathcal{X}(x_a)$ and $\mu:\mbox{Im}\,V' \rightarrow
\mathbb{R}_+$ assigns a nonnegative real number to every element in
the image of $V'$. For the proof of Lemma~\ref{lem:moreau_lyapunov},
refer to that of \cite[Theorem~4]{Mor:05}; see also
\cite[Section~4.5]{Khalil:02}. In the sequel, we will construct two
functions that satisfy the conditions in
Lemma~\ref{lem:moreau_lyapunov}.

First consider $V(x,s)$, $(x,s) \in \mathcal{X}(x_a)$ in
(\ref{eq:statespace}), given by
\begin{align} \label{eq:lyafcn}
V(x,s) := \frac{\textbf{1}^T (x+s)}{n} - \underline{m}(x).
\end{align}
Clearly $V$ depends continuously on $(x,s)$. {\color{black} Take any
finite $(x,s) \in \mathcal{X}(x_a)$; then both $\textbf{1}^T
(x+s)/n$ and $\underline{m}(x)$ are finite. Thus $V$ is bounded on
any bounded subsets of $\mathcal{X}(x_a)$.} Since $\textbf{1}^T
(x(k)+s(k))/n = \textbf{1}^T x(0)/n = x_a$ for all $k$, we obtain by
(ii), (iii) of Lemma~\ref{lem:property} that $V(x,s)$ is
non-increasing (i.e., $V(x(k_1),s(k_1)) \geq V(x(k_2),s(k_2))$ if
$k_1 \leq k_2$), and positive definite with respect to the average
consensus point $(x_a\textbf{1},0)$ (i.e., $V(x_a\textbf{1},0)=0$
and $V(x,s)>0$ if $(x,s) \neq (x_a\textbf{1},0)$).

Second, for a given $\kappa$ let $\delta_{\kappa}(x,s)$, $(x,s) \in
\mathcal{X}(x_a)$ in (\ref{eq:statespace}), be {\color{black}
\begin{align} \label{eq:posdeffcn} \delta_{\kappa}(x,s) :=
&\inf_{\zeta_0,\zeta_1,\ldots,\zeta_\kappa} V(\zeta_0) -
V(\zeta_\kappa),
\end{align}
where the infimum is taken over all sequences
$\zeta_0,\zeta_1,\ldots,\zeta_\kappa \in \mathcal{X}(x_a)$
satisfying
\begin{align*}
\zeta_0 &= (x,s) \\
\zeta_1 &= M(k)\zeta_0 \\
&\ \ \vdots \\
\zeta_\kappa &= M(k+\kappa-1)\zeta_{\kappa-1}
\end{align*}
for a given $k$. Thus $\zeta_i$, $i\in[1,\kappa]$, are the pairs of
states and surpluses possibly reachable from $(x,s)$ in $i$ time
steps.}


{\color{black} \begin{lem} \label{lem:delta_conti} The function
$\delta_{\kappa}:\mathcal{X}(x_a) \rightarrow \mathbb{R}_+$ in
(\ref{eq:posdeffcn}) is continuous in $(x,s)\in \mathcal{X}(x_a)$.
\end{lem}}
{\em Proof.} For given $k,\ \kappa $, consider an arbitrary sequence
$(x(k),s(k)),(x(k+1),s(k+1)),\ldots,(x(k+\kappa),s(k+\kappa))$
satisfying
\begin{align*}
&\begin{bmatrix}
  x(k+1)\\
  s(k+1)
\end{bmatrix}=
M(k)
\begin{bmatrix}
  x(k)\\
  s(k)
\end{bmatrix},\ \ldots,\
\\ & \begin{bmatrix}
  x(k+\kappa)\\
  s(k+\kappa)
\end{bmatrix}
=M(k+\kappa-1)
\begin{bmatrix}
  x(k+\kappa-1)\\
  s(k+\kappa-1)
\end{bmatrix}.
\end{align*}
First, we show that each $M(l)$, $l=k,\ldots,k+\kappa$, is a
continuous function of $(x,s)$. {\color{black} According to
(\ref{eq:state}) and (\ref{eq:surplus}), it suffices to show that
each of the functions $x_i:\mathbb{R}^{2n} \rightarrow \mathbb{R}$
and $s_i:\mathbb{R}^{2n} \rightarrow \mathbb{R}$, $i \in
\mathcal{V}$, is continuous in $(x,s)$. For this, let $y_i :=
\sum_{j \in \mathcal {N}_i^+} a_{ij} (x_j-x_i)$ and
$f\left(y_i\right) := c_i y_i$.  By (P4)
\begin{align*}
f\left(y_i\right) = \left\{
                         \begin{array}{ll}
                           y_i, & \hbox{$y_i \leq 0$;} \\
                           0, & \hbox{$y_i>0$.}
                         \end{array}
                       \right.
\end{align*}
Clearly $f$ is continuous in $y_i$. Since $y_i$ is a linear function
of $x$, function $f$ is continuous in $x$. Now substituting the term
$(x_i(k+1)-x_i(k))$ from (\ref{eq:state}) into (\ref{eq:surplus}),
we derive that $s_i$ is continuous in $(x,s)$. It then follows from
(\ref{eq:state}) that $x_i$ is also continuous in $(x,s)$.}


Second, the sequence
$(x(k),s(k)),(x(k+1),s(k+1)),\ldots,(x(k+\kappa),s(k+\kappa))$
depends continuously on $(x(k),s(k))$.  This is because each
function $M(l)$, $l=k,\ldots,k+\kappa$, is continuous, and there is
only a finite number of possible switching sequences of $\kappa-1$
digraphs. Thus, it follows from (\ref{eq:lyafcn}) that the
expression $V(x(k),s(k)) - V(x(k+\kappa),s(k+\kappa))$ depends
continuously on $(x(k),s(k))$. Finally, by the infimum definition of
(\ref{eq:posdeffcn}), we conclude that the function
$\delta_{\kappa}(x,s)$ is continuous in $(x(k),s(k))$. \hfill
$\blacksquare$

Now from (\ref{eq:posdeffcn}), one may easily see that the function
$\delta_{\kappa}(x,s)=0$ if $V(x,s)=0$; so
$\delta_{\kappa}(x_a\textbf{1},0)=0$.  The following result will be
vital, which asserts that there always exists a finite $\kappa $
such that the function $\delta_{\kappa}(x,s)$ is positive definite
with respect to the average consensus point $(x_a\textbf{1},0)$,
provided that the digraph is jointly strongly connected.

\begin{lem} \label{lem:posdeffcn}
Suppose that the dynamic digraph $\mathcal {G}(k)$ is jointly
strongly connected. There exists a finite $\kappa $ such that if
$V(x,s)$ is strictly positive, then $\delta_{\kappa}(x,s)$ is also
strictly positive.
\end{lem}

{\color{black} In the proof of Lemma~\ref{lem:posdeffcn}, we will
derive that a valid $\kappa$ value is $\kappa = (n-1)(n+1)\mathcal
{K}$, where $\mathcal {K}$ is the period when the dynamic network is
jointly strongly connected.  This $\kappa$ value is the one we will
use in the proof of Theorem~\ref{thm:dyn_alg}.}
Lemma~\ref{lem:posdeffcn} indicates that the function satisfies
$\delta_{\kappa}(x,s)>0$ for $(x,s) \neq (x_a\textbf{1},0)$. We
postpone the proof of Lemma~\ref{lem:posdeffcn}, and provide now the
proof of Theorem~\ref{thm:dyn_alg}.

{\em Proof of Theorem~\ref{thm:dyn_alg}.} (Sufficiency) Suppose that
$\mathcal {G}(k)$ is jointly strongly connected. Then it follows
from Lemmas~\ref{lem:delta_conti} and \ref{lem:posdeffcn} that the
function $\delta_{\kappa}$ defined in (\ref{eq:posdeffcn}) and the
function $V$ defined in (\ref{eq:lyafcn}) satisfy the conditions in
Lemma~\ref{lem:moreau_lyapunov}. Therefore uniform average consensus
is achieved.

(Necessity) Suppose that $\mathcal {G}(k)$ is not jointly strongly
connected. Namely for every $\mathcal {K}$ there exists $k_0$ such
that the union digraph $\mathcal{G}([k_0,k_0+\mathcal{K}])$ is not
strongly connected.  Thus during this interval $[k_0,k_0+\mathcal
{K}]$, there are some nodes not globally reachable; denote the
number by $r \in [1,n]$.  Case 1: $r=n$ (i.e., there is no globally
reachable node). Then $\mathcal{G}([k_0,k_0+\mathcal{K}])$ has at
least two distinct closed strong components, say $\mathcal {V}_1$
with $n_1$ nodes and $\mathcal {V}_2$ with $n_2$ nodes such that
$n_1+n_2=n$ (by \cite[Theorem~2.1]{Lin:Mono:08}). Consider a
state-surplus pair $(x(k_0),s(k_0))$ such that the nodes in
$\mathcal {V}_1$ have states $a$, those in $\mathcal {V}_2$ have
states $b$, and $a \neq b$; all surpluses are zero, $s(k_0)=0$. In
this case, no update of state or surplus will occur. One computes
that $|| (x(k_0),s(k_0)) - (x_a\textbf{1}, 0) ||_\infty =
\max\{|(a-b)n_2/n|, |(b-a)n_1/n|\}$; let $c_2=|| (x(k_0),s(k_0)) -
(x_a\textbf{1}, 0) ||_\infty$ and $c_1=c_2+\lambda$, $\lambda > 0$.
Then $|| (x(k_0),s(k_0)) - (x_a\textbf{1}, 0) ||_\infty < c_1$ but
$|| (x(k_0+\mathcal {K}),s(k_0+\mathcal {K})) - (x_a\textbf{1}, 0)
||_\infty = c_2$. Therefore uniform average consensus is not
achieved.

Case 2: $r<n$. We denote by $\mathcal {V}_g$ the set of all globally
reachable nodes.  Then $\mathcal {V}_g$ is the unique closed strong
component in $\mathcal{G}([k_0,k_0+\mathcal{K}])$ (again by
\cite[Theorem~2.1]{Lin:Mono:08}).  Consider a state-surplus pair
$(x(k_0),s(k_0))$ such that the nodes in $\mathcal {V}_g$ have
states $a$, those in $\mathcal {V}-\mathcal {V}_g$ have states $b$,
and $a \neq b$; all surpluses are zero, $s(k_0)=0$.  In this case,
no update will occur for the states in $\mathcal {V}_g$. Let $c_1=||
(x(k_0),s(k_0)) - (x_a\textbf{1}, 0) ||_\infty +\lambda$, $\lambda >
0$, and $c_2=|a-x_a|=|(a-b)(n-r)/n|$. Then $|| (x(k_0),s(k_0)) -
(x_a\textbf{1}, 0) ||_\infty < c_1$ but $|| (x(k_0+\mathcal
{K}),s(k_0+\mathcal {K})) - (x_a\textbf{1}, 0) ||_\infty \geq c_2$.
Therefore uniform average consensus is not achieved.~\hfill
$\blacksquare$

In the necessity proof above, Case 2 shows that even if $\mathcal
{G}(k)$ jointly contains a globally reachable node (e.g.,
\cite{Mor:05,RenBea:08,Lin:Mono:08}), uniform average consensus
cannot be achieved for certain state and surplus conditions. In
fact, state averaging requires the stronger connectivity notion:
jointly strongly connected $\mathcal {G}(k)$.

Finally we prove Lemma~\ref{lem:posdeffcn}. {\color{black} By the
definitions of $\delta_{\kappa}$ in (\ref{eq:posdeffcn}) and $V$ in
(\ref{eq:lyafcn}), it must be shown that there exists a finite
$\kappa$ such that for every time $k_0$ the minimum state satisfies
$\underline{m}(x(k_0)) < \underline{m}(x(k_0+\kappa))$. The proof is
organized into two steps. First we show that if some nodes have
positive surpluses, then all nodes in the network will have positive
surpluses after a finite time. Second, we show that using positive
surpluses, the nodes having the minimum state will increase their
values after a finite time. Although some other nodes may decrease
their state values, it is justified that the minimum state of the
whole network increases. The proof relies mainly on the graphical
condition of jointly strong connectedness as well as the state and
surplus update dynamics (\ref{eq:state}) and (\ref{eq:surplus}).}

{\em Proof of Lemma~\ref{lem:posdeffcn}.} Fix an arbitrary time $k_0
$, and denote by $\mu:=\underline{m}(x(k_0))$ the minimum state at
this time. Assume $\mu < x_a$ (i.e., average consensus is not yet
reached); thus $V(x(k_0),s(k_0))$ is strictly positive. It must be
shown that $\delta_{\kappa}(x(k_0),s(k_0))$ is also strictly
positive, for some finite $\kappa$.  This amounts to, by the
definitions of $\delta_{\kappa}$ in (\ref{eq:posdeffcn}) and $V$ in
(\ref{eq:lyafcn}), showing that $\mu <
\underline{m}(x(k_0+\kappa))$. {\color{black} We proceed in two
steps.

Step 1: We prove the following claim, which asserts that positive
surpluses can diffuse across the network under jointly strongly
connected topology.}

\emph{Claim.} Suppose that at time $k \geq k_0$ there are $r \in
[1,n-1]$ surpluses strictly positive, say $s_1(k), \ldots,
s_r(k)>0$, and $s_{r+1}(k)=\cdots=s_n(k)=0$. Then $s_i(k +
(n-r)\mathcal {K})>0$, for every $i \in \mathcal {V}$.

To prove the claim, we introduce a set $\mathcal{B}(k)$, $k \geq
k_0$, given by
\begin{align} \label{eq:possur}
\mathcal{B}(k) := \{i \in \mathcal {V} : s_i(k)>0 \}.
\end{align}
By the assumption of the claim, $\mathcal{B}(k)$ is a proper subset
of $\mathcal {V}$ (namely, $\mathcal{B}(k)\neq\emptyset, \mathcal
{V}$). First, owing to the surplus update (\ref{eq:surplus}),
together with (P1) and (P3), any strictly positive surplus cannot
decay to zero in finite time. This indicates $\mathcal{B}(k)
\subseteq \mathcal{B}(k+1)$, $k \geq k_0$. Next, since $\mathcal
{G}(k)$ is jointly strongly connected, there is an instant $\bar{k}$
in the interval $[k,k+\mathcal {K}]$ such that a directed edge
$(h,j)$ exists, for some $h \in \mathcal{B}(\bar{k})$ and some $j
\in \mathcal {V}-\mathcal{B}(\bar{k})$. Then agent $j$ receives
surplus of the amount $b_{ij}(\bar{k}) s_i(\bar{k})>0$, and hence
$\mathcal{B}(k)$ is strictly contained in $\mathcal{B}(k+\mathcal
{K})$. Repeating this argument leads to the conclusion that
$\mathcal{B}(k + (n-r)\mathcal {K})=\mathcal {V}$, which shows the
claim.

{\color{black} Step 2: Applying the above claim, we establish that
the minimum state of the network increases after a finite time
$\kappa$.} To this end, let another set $\mathcal{A}(k)$, $k \geq
k_0$, be
\begin{align} \label{eq:minset}
\mathcal{A}(k) := \{i \in \mathcal {V} : x_i(k)=\mu \}.
\end{align}
Then $\mathcal{A}(k)$ is the set of agents whose states are equal to
$\mu$ at time $k \geq k_0$.  First, owing to the state update
(\ref{eq:state}), together with (P2) and (P4), any $x_j(k)>\mu$
cannot decrease to $\mu$ in finite time.  This implies
$\mathcal{A}(k+1) \subseteq \mathcal{A}(k)$, $k \geq k_0$. Next, we
will establish that when the topology $\mathcal {G}(k)$ is jointly
strongly connected of period $\mathcal {K}$, there exists
$\tilde{\kappa}(\mathcal {K}) \in \mathbb{Z}_+$ such that
$\mathcal{A}(k+\tilde{\kappa}(\mathcal {K}))$ is strictly contained
in $\mathcal{A}(k)$, $k \geq k_0$ (that is,
$\mathcal{A}(k+\tilde{\kappa}(\mathcal {K}))$ has strictly less
agents than $\mathcal{A}(k)$).

We distinguish three cases. (i) $\mathcal{B}(k)=\mathcal {V}$. Under
jointly strongly connected topology, there is a directed edge
$(h,j)$, $h \in \mathcal {V}-\mathcal{A}(\bar{k})$ and $j \in
\mathcal{A}(\bar{k})$, for some time $\bar{k} \in [k,k+\mathcal
{K}]$. Then by (\ref{eq:state}) and (P4) we have
$x_j(\bar{k}+1)=x_j(\bar{k})+\epsilon_j(\bar{k})s_j(\bar{k})>x_j(\bar{k})\geq
x_j(k)$. So $\mathcal{A}(k+\mathcal {K})$ is strictly contained in
$\mathcal{A}(k)$. (ii) $\mathcal{B}(k)$ is a proper subset of
$\mathcal {V}$. It follows from the above claim that $\mathcal{B}(k
+ (n-r)\mathcal {K})=\mathcal {V}$.  Then by the same argument as in
case~(i) we obtain that $\mathcal{A}(k+(n-r+1)\mathcal {K})$ is
strictly contained in $\mathcal{A}(k)$. (iii)
$\mathcal{B}(k)=\emptyset$. Owing again to jointly strongly
connected topology, there is a directed edge $(h,j)$, with
$x_h(\bar{k}) < \overline{m}(k)$ and $x_j(\bar{k}) =
\overline{m}(k)$ (here $\overline{m}(k)$ is the maximum state at
time $k$), for some time $\bar{k} \in [k,k+\mathcal {K}]$. Then by
(\ref{eq:state}), (\ref{eq:surplus}), and (P4) we have
$s_j(\bar{k}+1)=-(x_j(\bar{k}+1)-x_j(\bar{k}))=-a_{jh}(\bar{k})(x_h(\bar{k})-x_j(\bar{k}))>0$,
and thereby $\mathcal{B}(k+\mathcal {K})=\{j\}$.  Now applying the
derivation in case~(ii) leads us to that
$\mathcal{A}(k+(n+1)\mathcal {K})$ is strictly contained in
$\mathcal{A}(k)$.  Summarizing the above three cases, and letting
$\tilde{\kappa}=(n+1)\mathcal {K}$, we obtain that
$\mathcal{A}(k+\tilde{\kappa})$ is strictly contained in
$\mathcal{A}(k)$.

Finally, since there are at most $n-1$ agents in $\mathcal{A}(k_0)$,
for $\kappa := (n-1)\tilde{\kappa}$ we have
$\mathcal{A}(k_0+\kappa)=\emptyset$. This implies $\mu <
\underline{m}(x(k_0+\kappa))$ with $\kappa = (n-1)(n+1)\mathcal
{K}$. \hfill $\blacksquare$


\section{Numerical Example} \label{Sec_Simul}

\begin{figure}[!t]
  \centering
  \includegraphics[width=0.7\textwidth]{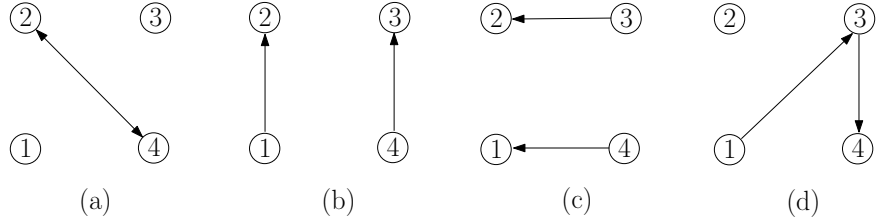}
  \caption{Periodically time-varying topology: a\ b\ c\ d\ a\ b\ c\ d\ $\cdots$.}
  \label{fig:Ex_Switch_4nodes}
\end{figure}

\begin{figure}[!t]
  \centering
  \includegraphics[width=0.65\textwidth]{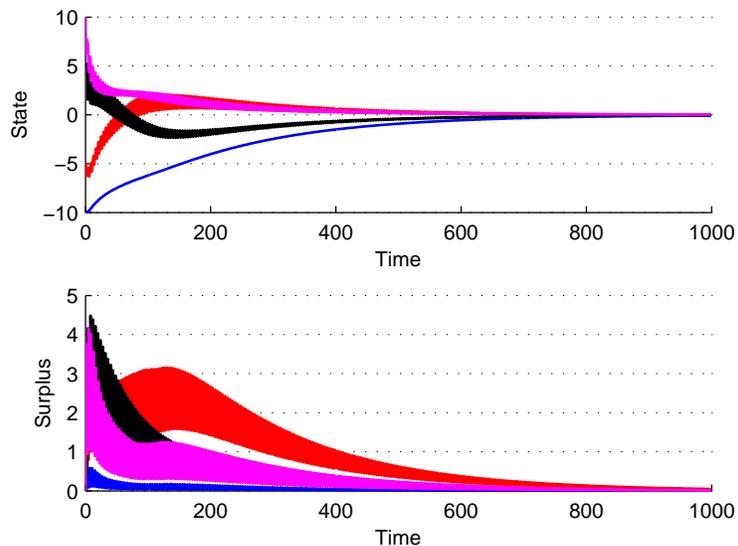}
  \caption{Convergence trajectories of states and surpluses obtained by applying the algorithm~(\ref{eq:dyn_alg}) for the topology in Fig.~\ref{fig:Ex_Switch_4nodes}}
  \label{fig:swi_exa}
\end{figure}

\begin{figure}[!t]
  \centering
  \includegraphics[width=0.65\textwidth]{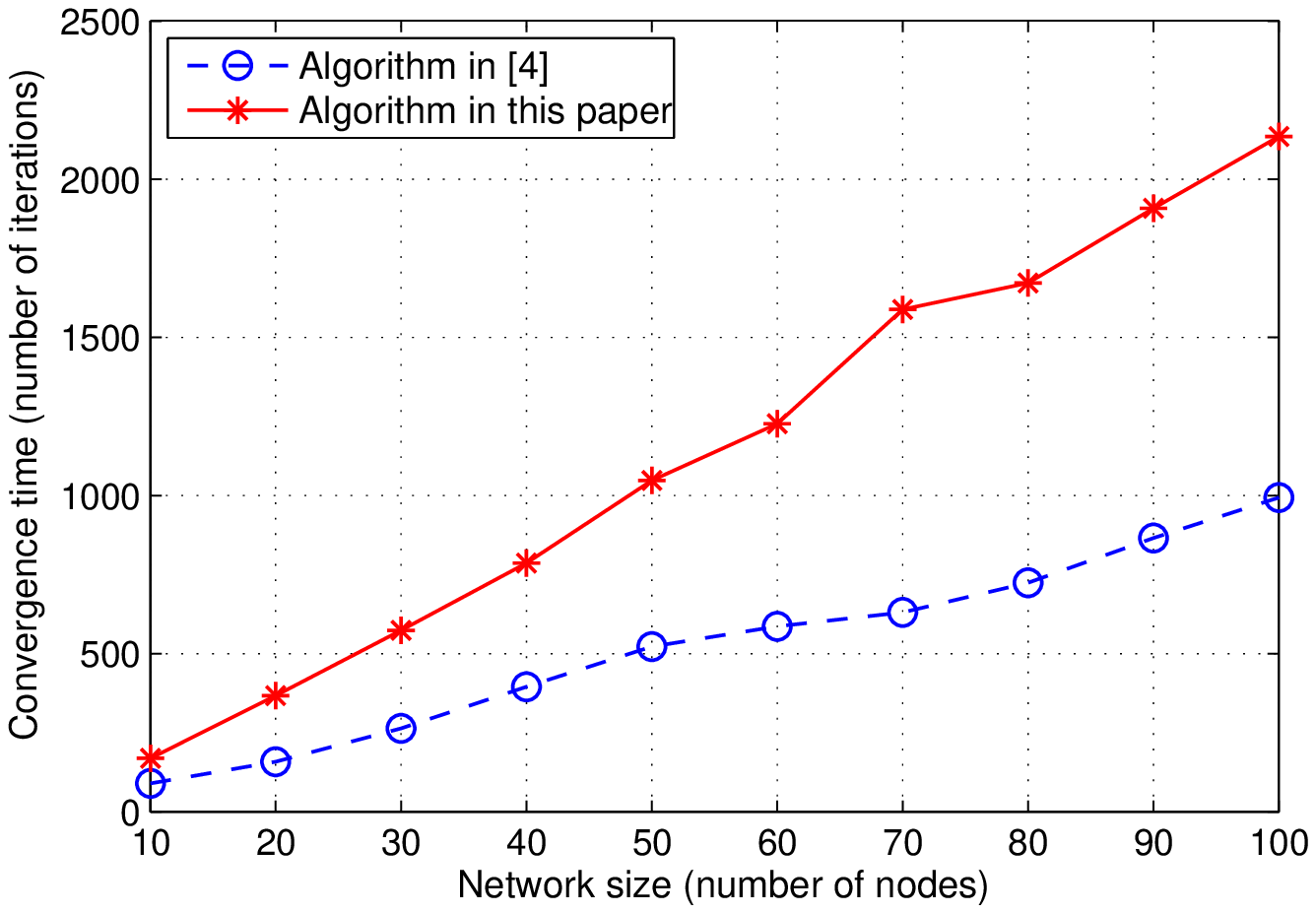}
  \caption{{\color{black} Convergence time comparison between the algorithm~(\ref{eq:dyn_alg}) and the one in \cite{CaiIshii_Aut:12}.
  Consider digraphs $\mathcal {G}_n$ with $n$ nodes derived from complete graphs with $n$ nodes by removing $n-2$ edges
  $(h,n)$, $h \in [2,n-1]$.  Thus $\mathcal {G}_n$ are strongly connected, non-balanced digraphs, and we consider static
  $\mathcal {G}_n$ since the algorithm in \cite{CaiIshii_Aut:12} applies only to this case.  For simplicity choose
  $a_{ij}=b_{ij}=1/n$ and $\epsilon_i=1/(2n)$. Run both algorithms on digraphs $\mathcal {G}_n$ with different $n$,
  and with initial states $x_i(0)$ randomly selected from $[-50,50]$,
  initial surpluses $s_i(0)=0$. Find the convergence times $t$ to be the minimum times when $|| (x(t),s(t))-(x_a\textbf{1},0) ||_1 < 0.05$, and
  each sample point on the displayed curves is the average convergence time of $50$ algorithm executions.  Comparing the two curves we see that
  the convergence time of algorithm~(\ref{eq:dyn_alg}) is approximately twice as much as the one in
  \cite{CaiIshii_Aut:12}, which is due to the former's enforcement
  on nonnegative surpluses that produces undesirable switchings. }}
  \label{fig:ConvTimeCompare}
\end{figure}

We provide a numerical example to illustrate the convergence result
of the algorithm~(\ref{eq:dyn_alg}). Consider the periodically
time-varying digraph $\mathcal {G}(k) = (\mathcal {V}, \mathcal
{E}(k))$, with period $\mathcal {K}=4$, displayed in
Fig.~\ref{fig:Ex_Switch_4nodes}. No single digraph is strongly
connected, but $\mathcal {G}(k)$ is jointly strongly connected. For
simplicity, we apply the algorithm~(\ref{eq:dyn_alg}) by choosing
the parameters and weights to be constant:
$\epsilon_h=a_{ij}=b_{ij}=1/4$ for all agents $h$ and all edges
$(j,i)$.  It is easily verified that this choice satisfies the
requirements (P1)-(P3).

For the initial state $x(0)=[-10\ \-5 \ 5\ 10]^T$ and the initial
surplus $s(0)=0$, the state and surplus trajectories are displayed
in Fig.~\ref{fig:swi_exa}. Observe that every state converges to the
desired average $0$, and every surplus is always nonnegative and
vanishes eventually.  Also we see that there are considerable
switchings in both states and surpluses due to the enforcement of
nonnegative surpluses, which may undesirably slow down the
convergence speed. {\color{black} Indeed, in a simulation study
displayed in Fig.~\ref{fig:ConvTimeCompare}, we compare the
algorithm~(\ref{eq:dyn_alg}) to the one in \cite{CaiIshii_Aut:12}
(the latter poses no restriction on nonnegative surpluses), and find
that the convergence time of algorithm~(\ref{eq:dyn_alg}) is
approximately twice as much as the one in \cite{CaiIshii_Aut:12} for
a class of digraphs.} An important future study then would be to
find appropriate (possibly time-varying) values of the parameters
and weights so as to reduce switchings and accelerate convergence.

\section{Conclusions} \label{Sec_Concl}

We have proposed a new surplus-based algorithm which enables
networks of agents to achieve uniform average consensus on general
time-varying digraphs that vary in some deterministic fashion. Our
derived graphical condition does not require balanced or symmetric
network topologies, and is hence more general than those previously
reported in the literature. Future research will target convergence
speed analysis of the algorithm as well as the design of fast
surplus-based averaging algorithms.


\bibliographystyle{IEEEtran}                   
\bibliography{DistributedControl,EigPurterb,SwitchedSystem}           

\begin{thebibliography}{10}
\providecommand{\url}[1]{#1}
\csname url@rmstyle\endcsname
\providecommand{\newblock}{\relax}
\providecommand{\bibinfo}[2]{#2}
\providecommand\BIBentrySTDinterwordspacing{\spaceskip=0pt\relax}
\providecommand\BIBentryALTinterwordstretchfactor{4}
\providecommand\BIBentryALTinterwordspacing{\spaceskip=\fontdimen2\font plus
\BIBentryALTinterwordstretchfactor\fontdimen3\font minus
  \fontdimen4\font\relax}
\providecommand\BIBforeignlanguage[2]{{%
\expandafter\ifx\csname l@#1\endcsname\relax
\typeout{** WARNING: IEEEtran.bst: No hyphenation pattern has been}%
\typeout{** loaded for the language `#1'. Using the pattern for}%
\typeout{** the default language instead.}%
\else
\language=\csname l@#1\endcsname
\fi
#2}}

\bibitem{BerTsi:89}
D.~P. Bertsekas and J.~N. Tsitsiklis, \emph{Parallel and Distributed
  Computation: Numerical Methods}.\hskip 1em plus 0.5em minus 0.4em\relax
  Prentice-Hall, 1989.

\bibitem{OlfMur:04}
R.~Olfati-Saber and R.~M. Murray, ``Consensus problems in networks of agents
  with switching topology and time-delays,'' \emph{IEEE Trans. Autom. Control},
  vol.~49, no.~9, pp. 1520--1533, 2004.

\bibitem{XiBo:04}
L.~Xiao and S.~Boyd, ``Fast linear iterations for distributed averaging,''
  \emph{Systems \& Control Letters}, vol.~53, pp. 65--78, 2004.

\bibitem{CaiIshii_Aut:12}
K.~Cai and H.~Ishii, ``Average consensus on general strongly connected
  digraphs,'' \emph{Automatica}, vol.~48, no.~11, pp. 2750--2761, 2012. Also in
  {\em Proc. 50th IEEE Conf. on Decision and Control and European Control
  Conf.}, pp. 1956-1961, 2011.

\bibitem{CaiIshii_ACC:10}
------, ``Quantized consensus and averaging on gossip digraphs,'' \emph{IEEE
  Trans. Autom. Control}, vol.~56, no.~9, pp. 2087--2100, 2011.

\bibitem{CaiIshii_ACC:12}
------, ``Average consensus on arbitrary strongly connected digraphs with
  dynamic topologies,'' in \emph{Proc. American Control Conf.}, no. 14-19,
  Montreal, Canada, 2012.

\bibitem{Mor:05}
L.~Moreau, ``Stability of multi-agent systems with time dependent communication
  links,'' \emph{IEEE Trans. Autom. Control}, vol.~50, no.~2, pp. 169--182,
  2005.

\bibitem{Lin:Mono:08}
Z.~Lin, \emph{Distributed {C}ontrol and {A}nalysis of {C}oupled {C}ell
  {S}ystems}.\hskip 1em plus 0.5em minus 0.4em\relax VDM Verlag, 2008.

\bibitem{HenTsi:13}
J.~M. Hendrickx and J.~N. Tsitsiklis, ``Convergence of type-symmetric and
  cut-balanced consensus seeking systems,'' \emph{IEEE Trans. Autom. Control},
  vol.~58, no.~1, pp. 214--218, 2013.

\bibitem{FraGiuSea:11}
M.~Franceschelli, A.~Giua, and C.~Seatzu, ``Distributed averaging in sensor
  networks based on broadcast gossip algorithms,'' \emph{IEEE Sensors J.},
  vol.~11, no.~3, pp. 808--817, 2011.

\bibitem{BBTTV:10}
F.~Benezit, V.~Blondel, P.~Thiran, J.~Tsitsiklis, and M.~Vetterli, ``Weighted
  gossip: distributed averaging using non-doubly stochastic matrices,'' in
  \emph{Proc. IEEE Int. Symposium on Information Theory}, Austin, TX, 2010, pp.
  1753--1757.

\bibitem{DomHad:13}
A.~D. Dominguez-Garcia and C.~N. Hadjicostis, ``Distributed matrix scaling and
  application to average consensus in directed graphs,'' \emph{IEEE Trans.
  Autom. Control}, no.~9, to appear, 2013. Also in {\em Proc. 50th IEEE Conf.
  on Decision and Control and European Control Conf.}, pp. 2124-2129, 2011.

\bibitem{GhaCor:11}
B.~Gharesifard and J.~Cort\'es, ``Distributed strategies for generating
  weight-balanced and doubly stochastic digraphs,'' \emph{European J. of
  Control}, vol.~18, no.~6, pp. 539--557, 2012. Also arXiv:0911.0232v4
  [math.OC].

\bibitem{RenBea:08}
W.~Ren and R.~W. Beard, \emph{Distributed Consensus in Multi-vehicle
  Cooperative Control: Theory and Applications}.\hskip 1em plus 0.5em minus
  0.4em\relax Springer-Verlag, 2008.

\bibitem{LiuAndCaoMor:09}
J.~Liu, B.~D.~O. Anderson, M.~Cao, and A.~S. Morse, ``Analysis of accelerated
  gossip algorithms,'' in \emph{Proc. 48th IEEE Conf. on Decision and Control},
  Shanghai, China, 2009, pp. 871--876.

\bibitem{Khalil:02}
H.~K. Khalil, \emph{Nonlinear Systems}.\hskip 1em plus 0.5em minus 0.4em\relax
  3rd ed., Prentice-Hall, 2002.

\end{thebibliography}

\end{document}